\documentclass[fleqn,usenatbib]{mnras}

\usepackage[T1]{fontenc}

\DeclareRobustCommand{\VAN}[3]{#2}
\let\VANthebibliography\thebibliography
\def\thebibliography{\DeclareRobustCommand{\VAN}[3]{##3}\VANthebibliography}

\usepackage{graphicx}	
\usepackage{subfig}
\usepackage{amsmath}	
\usepackage{amssymb}	
\usepackage{multicol}        
\usepackage{multirow}
\usepackage{hyperref}
\usepackage{academicons}
\usepackage{threeparttable}
\usepackage{soul,xcolor}
\usepackage[normalem]{ulem}
\definecolor{orcidlogocol}{HTML}{A6CE39}
\graphicspath{{./}{figures/}}

\title[Bar formation in submaximal discs]{Bar formation in submaximal discs: a challenge for $\Lambda$CDM}

\author[T. Kashfi et al.]{
Tahere Kashfi,$^{1}$
Mahmood Roshan,$^{1,2}$\thanks{E-mail: mroshan@um.ac.ir}
Benoit Famaey$^{3}$\thanks{E-mail: benoit.famaey@astro.unistra.fr}
\\
$^{1}$Ferdowsi University of Mashhad, P.O. Box 1436, Mashhad, Iran\\
$^{2}$Institute for Research in Fundamental Sciences (IPM), 19395-5531, Tehran, Iran\\
$^{3}$Universit\'e de Strasbourg, CNRS, Observatoire astronomique de Strasbourg, UMR 7550, F-67000 Strasbourg, France
}

\date{Accepted XXX. Received YYY; in original form ZZZ}

\pubyear{2022}

\begin{document}
\label{firstpage}
\pagerange{\pageref{firstpage}--\pageref{lastpage}}
\maketitle

\begin{abstract}
Bar formation in cosmological simulations of galaxy formation remains challenging. It was previously shown that the fraction of barred galaxies at low stellar masses ($M_*<10^{10.5} M_\odot$) in TNG50 is too low compared to observations. Here, we highlight another tension, also observed at higher stellar masses, namely that barred galaxies in TNG50 appear to be maximal discs, in the sense that the majority of the gravitational acceleration is accounted for by baryons at the peak radius of the baryonic rotation curve, while observations account for a substantial fraction of barred submaximal discs. In this letter, we compare the barred fraction of submaximal discs in the local Universe from the SPARC catalogue with that in the TNG50 simulation. We show that, although SPARC tends to select against barred galaxies, the fraction of barred submaximal discs in this dataset is significantly larger than in TNG50. This result adds to the list of challenges related to predicting the right statistics and properties of barred galaxies in $\Lambda$CDM simulations of galaxy formation.
\end{abstract}




\section{Introduction}
The rotation curve decomposition of spiral galaxies is a primary tool for measuring the contribution of baryonic and dark matter components. However, without a prior on the mass-to-light ratio of the stellar component, it is difficult to determine the corresponding contributions of each component because of the underlying degeneracy in mass-decomposition. This problem led to the maximum disc hypothesis \citep{Albada1986} which sets a lower limit on the dark matter content in the galaxy. According to the standard definition, galactic discs are maximal if the disc contribution in the total rotation curve at $R\approx 2.2\,R_d$ is $0.75 \leq V_{\text{disc}}/V_{\text{tot}} \leq 0.95$ \citep{Sackett1997}, where $R_d$ is the scale-length of the disc. This definition can be generalized at the peak radius of the baryonic rotation curve to include gas-dominated discs. Modern stellar population synthesis models \citep[e.g.][]{Schombert2019} tend to back this hypothesis in most high surface brightness disc galaxies: some controversial studies indicate that most spiral galaxies could be submaximal \citep{Bershady2011, Martinsson2013}, however, see \citet{Aniyan2018,Aniyan2021} for a rebuttal of such claims. 

What is, on the other hand, certain is that not all disc galaxies are maximal: this is particularly not the case in low surface brightness galaxies that tend to be dark matter dominated all the way down to the centre \citep[e.g.,][]{McGaugh1998}. Because of the dark matter dominated nature of submaximal discs, they are a golden place to explore the role of the dark matter component in the evolution of galaxies.

One of the most important processes in the evolution of spiral galaxies is the bar formation which is directly influenced by the presence of the dark matter halo. For instance, the exponential growth rate of a bar in a close-to-maximal idealized stellar disc is substantially larger when the disc is embedded in a live dark matter halo than in a rigid one having the same mass distribution \citep{Sellwood2016}. Things become however more complicated in a cosmological context, where it has been shown that bar growth tends to be fully \citep{Reddish2022} or partially suppressed, with no bar at the low mass end and too short bars at the high mass end \citep[or too slow bars for a given size,][]{Roshan2021, Frankel2022}, although see \citet{Fragkoudi2021} for a simulation producing realistic bars at the high mass/high surface brightness end. On the other hand, in submaximal discs, a large amount of dark matter content in the central parts stabilizes the disc against the bar instability both in idealized and cosmological contexts \citep[e.g.,][]{Mihos1997, DeBuhr2012, Yurin2015, Algorry2017}. However, it is conceivable that some mechanisms beyond the secular evolution of the disc within its live dark matter halo could trigger bars in submaximal discs in a cosmological context, such as tidal interaction with a close neighbour \citep{Izquierdo2022} or very efficient cooling in the gaseous disc \citep[e.g.][]{Mayer2004}. This should be especially true in formally submaximal but `near-maximal' discs.

The main purpose of the present short letter is to study the barred submaximal discs statistics in the Illustris TNG50 cosmological simulation \citep{Nelson2019a, Nelson2019, Pillepich2019}, where the above mechanisms should naturally take place, and compare these statistics to observational ones in the local Universe. In \citet{Roshan2021}, we have already shown that the bar fraction in TNG50 was increasing sharply from effectively zero at stellar masses of $\sim 10^{10} M_\odot$ to almost 90\% at $\sim 10^{11} M_\odot$, in apparent contradiction with the results of \citet{Erwin2018} derived from the Spitzer Survey of Stellar Structure in Galaxies (S$^4$G) of the local Universe, where the observed bar fraction rather peaks at a stellar mass of $10^{9.7} M_\odot$. The lack of barred galaxies remains clear for $M_* < 10^{10.5} M_\odot$ in TNG50 even when comparing with other observational bar fraction estimates \citep{Oh2012}. Note that the measured bar fraction in TNG50 at $M_* < 10^{10.5} M_\odot$ is, however, higher in \citet{Zana2022} than in \citet{Roshan2021}, but still not quite as high as in \citet{Oh2012}. Note that \citet{Masters2012, Melvin2014} found even smaller observational bar fractions at small masses, as well as an increasing bar fraction with mass in contradistinction with various other studies \citep{Diaz2016, Consolandi2016}.

Notwithstanding this possible deficit of barred galaxies with $M_* < 10^{10.5} M_\odot$ in TNG50, we focus our present study on submaximal discs, including galaxies with higher stellar masses for which the fraction of barred galaxies in TNG50 is substantial. We will compare the fraction of simulated barred submaximal discs to that in an observational sample biased against barred galaxies, meaning that a lower fraction in simulations would indicate a potential problem. 

The structure of this letter is as follows: In section \ref{sec:SPARC} we study the submaximal discs in an observational sample. In section \ref{sec:TNG50}, we perform our main analysis within the TNG50 cosmological simulation. Final comments and conclusions can be found in section \ref{sec:conc}.
\section{Submaximal discs in observations of massive spirals} \label{sec:SPARC}
We employ the \textit{Spitzer} Photometry and Accurate Rotation Curves (SPARC) database \citep{Lelli2016}, including 175 galaxies. SPARC is the largest galaxy sample which includes near-infrared ($3.6\mu m$) surface brightness and high-quality $HI/H\alpha$ rotation curves. SPARC galaxies are representative of the disc population in the field and in nearby groups like Ursa Major \citep{Lelli2016}. This sample spans the broadest range of disc galaxy properties in the field in terms of luminosity, stellar mass, surface brightness, and morphological type, but it explicitly biases against barred galaxies, as the sample is chosen for high-quality rotation curves with not too important non-axisymmetric motions. This means that the fraction of barred galaxies in this sample is not representative but can be taken as a {\it lower limit} on the true bar fraction.

To ensure a fair comparison with TNG50, we will restrict ourselves to galaxies with stellar masses $M_* \geq 10^{10} M_\odot$ in SPARC, i.e. to a total of 59 galaxies. In this mass range, most spiral galaxies are high-surface brightness maximal discs; however, a few submaximal discs exist too. As mentioned before, comparing the disc and observed rotation curve at $R\approx 2.2 R_d$ can provide a reasonable estimate of the disc maximality because for a pure exponential disc, the peak of the baryonic rotation curve occurs at this radius. However, for many galaxies, the baryonic distribution deviates from an exponential profile (such as in bulge-dominated galaxies), or the gas contribution entirely dominates the baryonic budget (such as in some dwarf galaxies). Thus, it is preferable to compare the baryonic and observed velocities at the radius where the total baryonic rotation curve peaks \citep{Martinsson2013, Starkman2018}.

In this work, we compute the baryonic maximality (stars+gas) to identify the submaximal discs. To do so, we set 
\begin{equation}
\mathcal{F}_{b}=V_{\text{bary}}^{\text{max}}(R_p)/V_{\text{tot}}(R_p)<0.75,
\label{criterion}
\end{equation} 
where the lower end of the maximality is considered \citep{Kregel2005}. Here, $V_{\text{bary}}^{\text{max}}$ is the maximum baryonic rotational velocity and $V_{\text{tot}}$ is the total rotation curve at the radius $R_p$ where the baryonic rotation curve peaks. In terms of gravitational acceleration, our definition of submaximality means that the fraction of gravity accounted for by baryons at the peak radius $R_p$ of the baryonic rotation curve is smaller than $\mathcal{F}_{b}^2 \sim 56\%$.

To ensure that we do not deal hereafter with artificially submaximal discs due to a misestimated inclination, we implemented the corrected rotation curves provided by \citet{Li2018} to investigate the bar fraction in massive submaximal discs in this sample. In \citet{Li2018}, individual galaxies are fitted to the radial acceleration relation (RAR) \citep{McGaugh2016} by imposing priors on $\Upsilon_{\star}$, galaxy distance ($D$) and disc inclination ($i$) based on the observational uncertainties. Then, using an MCMC method, the best fit value for these free parameters is found. Importantly, with this correction, the fraction of barred submaximal discs in SPARC is smaller than when not applying the correction. This ensures that we are being conservative and that the maximality of disc galaxies has not been underestimated.

\begin{figure}
	\centering
	\includegraphics[width=0.85\linewidth]{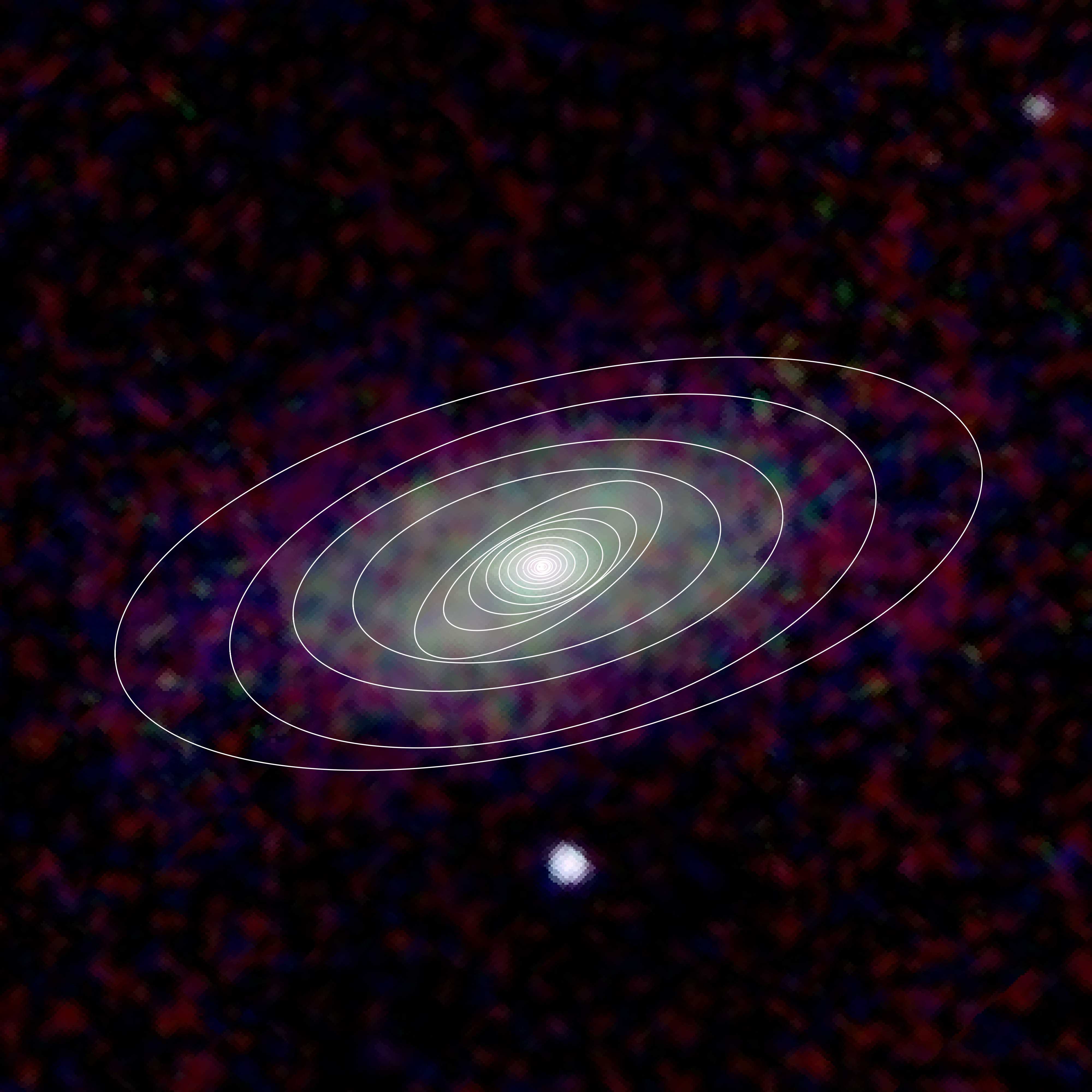}
	\caption{The image of galaxy NGC~1090 in the infrared band from 2MASS oriented north up and east left. The white ellipses are derived using the isophotal ellipse fitting method.\label{fig:NGC1090}}
\end{figure}

\begin{figure}
	\centering
	\includegraphics[width=0.95\linewidth]{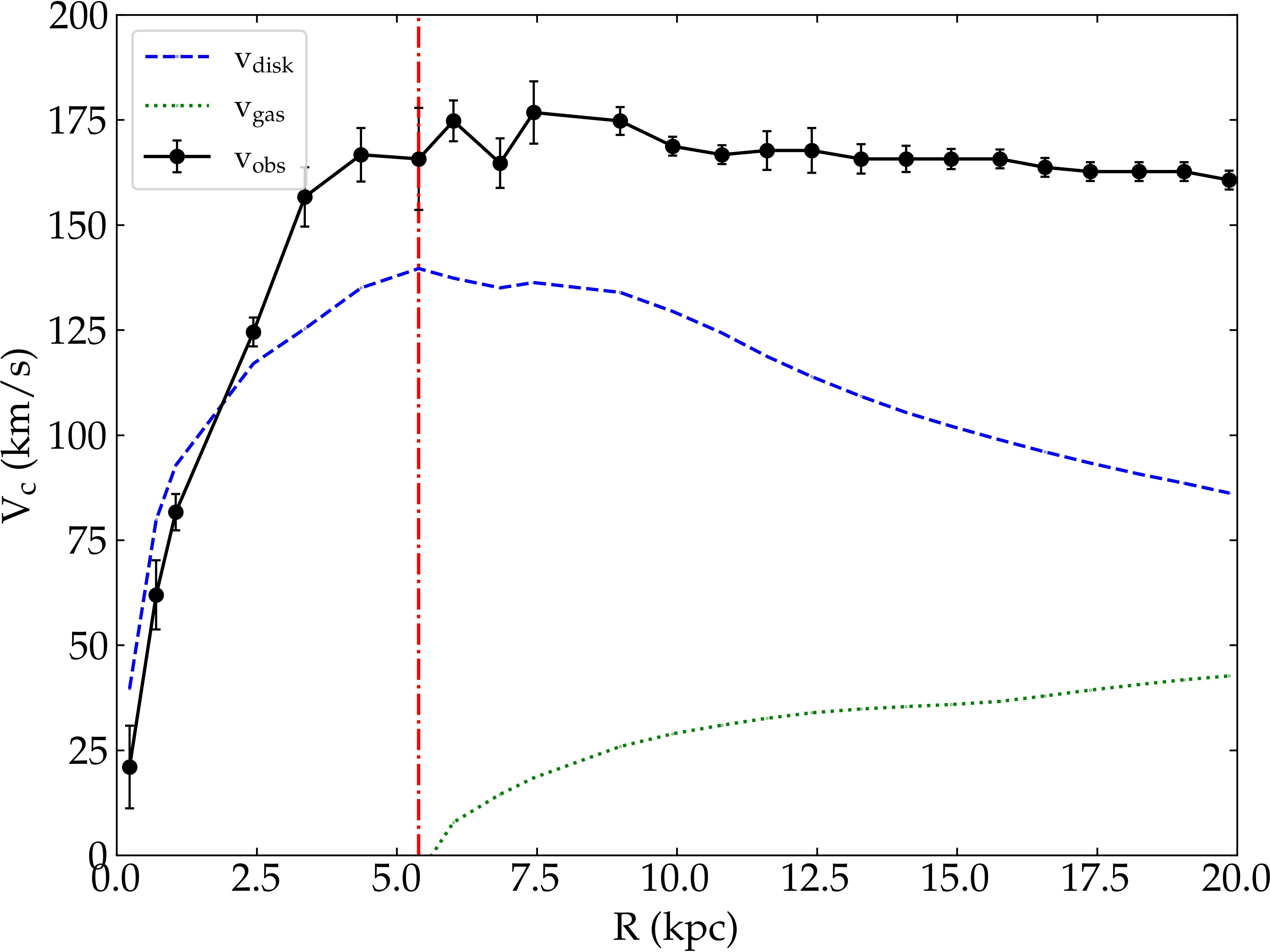}
	\caption{SPARC rotation curve of NGC~1090. The black points with error bars are the observed rotation curve $V_{obs}(R)$. Each baryonic component is presented: the blue dashed line for the disc and the green dotted line for the gas. This disc can be considered `maximal' in the sense that the stellar mass-to-light ratio cannot be higher, as evident from the inner rotation curve points, but it is slightly submaximal in the sense that the gravitational acceleration provided by the baryons at $R_p$ is only 52\% ($\mathcal{F}_{b}=0.72$).\label{fig:RC_NGC1090}}
\end{figure}

Among the 59 galaxies with $M_* \geq 10^{10} M_\odot$ in SPARC, we find  14 submaximal discs according to our definition in Eq.~1, i.e. a fraction of 23.73 \%. The majority of those are relatively close to the submaximality cut-off, meaning that they are mostly `near-maximal', which is expected at such stellar masses. 

Among these submaximal discs, we use the morphology types reported in the SPARC database to identify the barred galaxies. This is illustrated with one example in Fig.~\ref{fig:NGC1090}, where we show an infrared image of the galaxy NGC~1090 where the bar is clearly apparent. The ellipses are determined from the fit of the galaxy isophotes by ellipses \citep{Aguerri2009}, and the misalignment of the bar major axis with the major axis of the projected disc on the sky is clearly visible in the figure. The SPARC rotation curve of this galaxy is shown in Fig.~\ref{fig:RC_NGC1090}. Although this procedure is not perfect, it ensures that a lower limit on the bar fraction can be found since the SPARC sample biases against the selection of barred galaxies and since only bars that are visually obvious are selected. 
Even with such restrictions, we find that 36\% of the massive submaximal discs in SPARC are barred. Note that our definition of submaximality really pertains to the amount of missing mass at the peak of the baryonic rotation curve, not to the fact that a higher stellar mass-to-light ratio could have been used for the disc.

As mentioned before, we used the corrected rotation curves of the SPARC galaxies according to \citet{Li2018}. So each galaxy has a $\chi^2$ parameter which indicates the goodness of fitting the free parameters to the radial acceleration relation. By removing galaxies with poor $\chi^2$, i.e., $\chi^2>10$, from the SPARC data set, we found that the bar fraction of massive submaximal discs in this sample is reduced to 3/11=27\%. We also imposed the minimum precision of $10\%$ in the observational velocity ($\delta V_{\text{obs}}/V_{\text{obs}}<0.1$), and this last requirement does not change our final results. In summary, no other selection led to a fraction of barred submaximal discs among galaxies with $M_* \geq 10^{10} M_\odot$ in SPARC lower than 27\%.


\section{Submaximal discs in cosmological simulations} \label{sec:TNG50}
Here, we use the highest resolution run of the TNG project that, in principle, includes all the matter components and a considerable amount of physical mechanisms that can affect bar formation in galactic discs in the $\Lambda$ cold dark matter ($\Lambda$CDM) paradigm. To find the fraction of submaximal barred galaxies in this cosmological simulation, we use two criteria to identify disc galaxies: (i) $k_{rot} \geqslant 0.5$, where $k_{rot}$ is the fraction of stellar kinetic energy in ordered rotation \citep{Sales2010}; (ii) the morphological flatness criterion $F \leqslant 0.7$, where $F$ depends on the eigenvalues of the moment of inertia tensor of the galaxy \citep{Genel2015} (see \citealt{Roshan2021} for more details).

\begin{figure}
	\centering
	\includegraphics[width=0.9\linewidth]{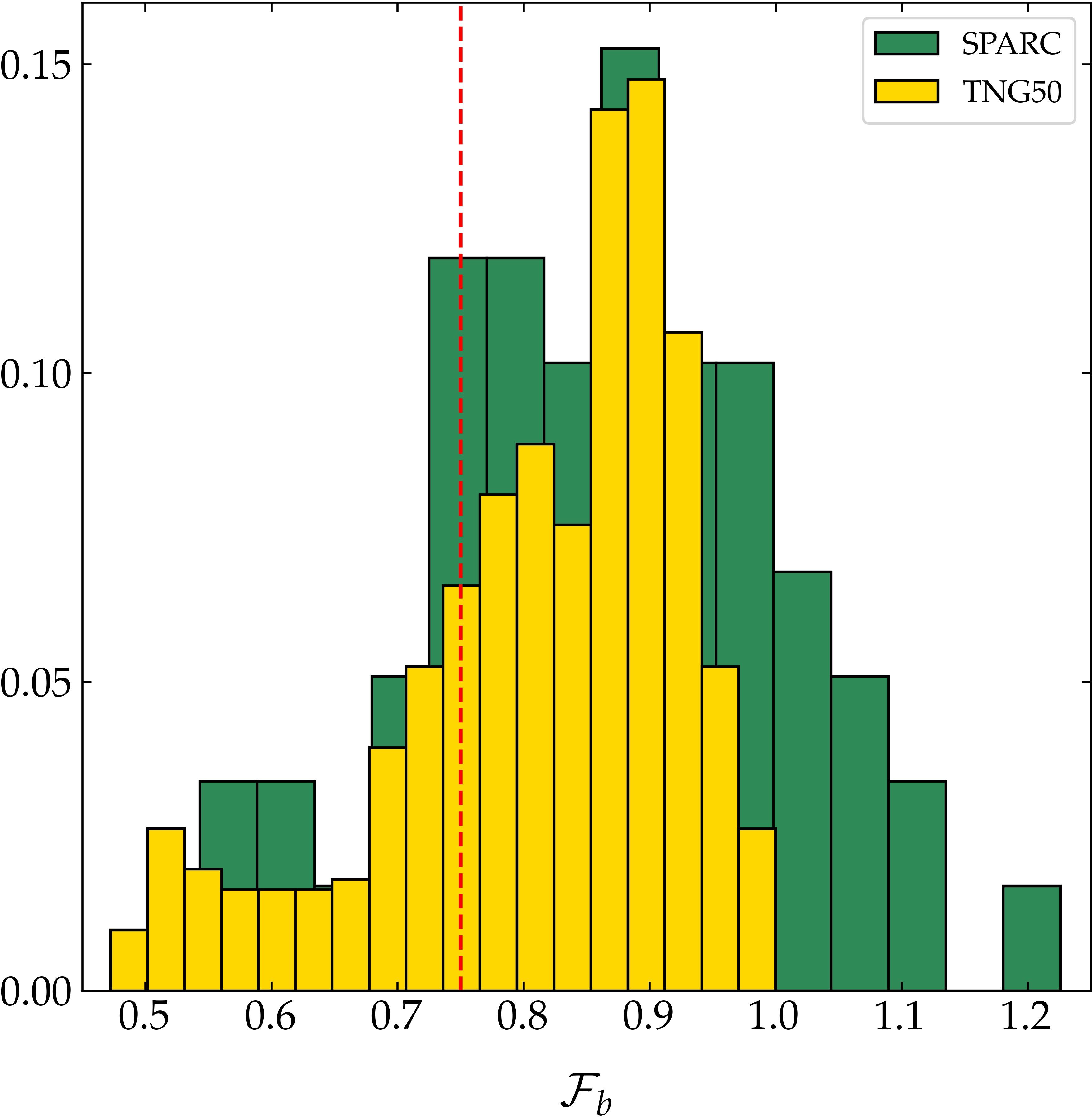}
	\caption{The distribution of disc galaxies with $M_{\star}\geq 10^{10} M_{\odot}$ in terms of baryonic maximality as defined in Eq.~\ref{criterion} for the SPARC sample (green bars) and TNG50 cosmological simulation (yellow bars). The vertical dashed line indicates the maximality cutoff.\label{fig:hist}}
\end{figure}

We implement a widely used relation for the rotation curves of each component, namely, $V_i(r)=\sqrt{GM_i(r)/r}$, where $M_i(r)$ denotes the total enclosed mass of each component at a given radius. The accuracy of using this approximation to compute the rotation curve is tested by \citet{Roshan2021}. They showed that the value of $V(r)$, which is derived by the total mass distribution, matches well with the rotation curve obtained based on the force calculation (see \citealt{Roshan2021}, figure 4). Only at the very central regions of the galaxy, using this approximation, has a lower accuracy. We have 610 disc galaxies in our sample with $M_{\star}\geq 10^{10} M_{\odot}$, where 155 of them are submaximal, i.e., $\mathcal{F}_b<0.75$. This fraction of 25.4\% of submaximal discs is comparable to the fraction of 23.73 \% in SPARC. The distribution in terms of the maximality parameter of Eq.~\ref{criterion} in SPARC and TNG50 is plotted in Fig.~\ref{fig:hist}.

To identify the fraction of barred submaximal discs in TNG50, we measure the bar strength, which is defined in terms of the Fourier decomposition of the surface density,
\begin{equation}\label{eq8}
A_2^{\text{max}}=\max[A_2(R)],
\end{equation}
Here $A_2(R)$ is the Fourier amplitude of mode $m=2$ at radius $R$. For arbitrary mode $m$, the amplitude is written in terms of Fourier coefficients
\begin{equation}\label{eq9}
A_m(R)\equiv \sqrt{a_m(R)^2+b_m(R)^2},
\end{equation}
where
\begin{align}\label{eq10}
&a_m(R)=\frac{1}{M(R)}\sum_{k=0}^{N}m_k\cos(m\phi_k),\  m=1,2,...\\
&b_m(R)=\frac{1}{M(R)}\sum_{k=0}^{N}m_k\sin(m\phi_k),\  m=1,2,...
\end{align}
To compute these coefficients, we divide the disc into annuli of equal width and consider only particles with $|z|<1$ kpc. In these equations, $N$ represents the number of particles in each annulus, and $M$ represents the total mass of the annulus with a mean cylindrical radius $R$. $A_2^{\text{max}}\geq 0.4$ indicates the strong bars and the unbarred bars have $A_2^{\text{max}}<0.2$. 

We find that only 5.8\% of the submaximal discs in TNG50 host a (weak) bar. Although the distribution of galaxies in terms of their maximality is not so different for these two data sets, as mentioned above (25.4\% vs 23.73\%), there is an obvious tension between the fraction of submaximal barred galaxies. If the true probability of a barred submaximal disc is given by TNG50 in the considered stellar mass range, the probability of having at least 5/14 barred submaximal discs in SPARC, as observed, is 0.085\% as per the binomial distribution. If we remove galaxies that deviate from the radial acceleration relation, which could indicate that they are perturbed, the bar fraction of massive submaximal discs is reduced to 3/11 = 27\%, and the probability increases to 2.26\% as per the binomial distribution.

To refine our comparison, we also explored the bar fraction in near-maximal discs: to do so; we chose galaxies whose maximalities are in the range of $0.7 < \mathcal{F}_{b} < 0.75$. We find that SPARC contains nine near-maximal galaxies, of which four of them are barred. On the other hand, TNG50 has 67 galaxies with $0.7 < \mathcal{F}_b < 0.75$ and 10\% of them host bars. If we consider the bar fraction in TNG50 as truth, by the binomial distribution, one can show that the probability of having four barred galaxies out of nine by chance is then about 0.8\%, still a very significant disagreement. 

\begin{figure}
	\centering
	\includegraphics[width=1.0\linewidth]{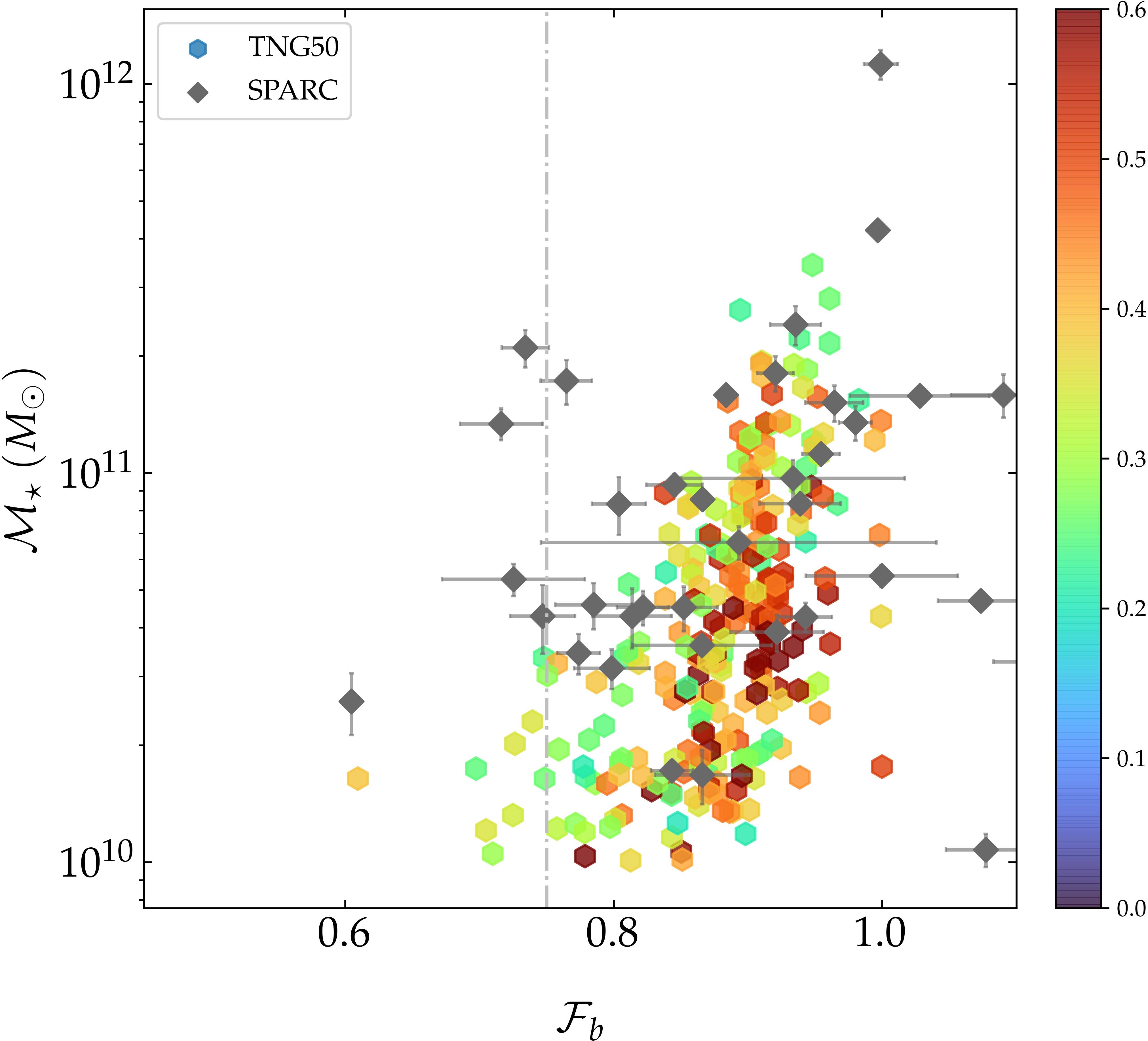}
	\caption{Stellar mass versus baryonic maximality for all barred galaxies in SPARC and TNG50 data sets. The colour shows the bar strength $A_2^{\text{max}}$ for galaxies in TNG50. \label{fig:barred-galaxies}}
\end{figure}

To visualize the disagreement, Fig.~\ref{fig:barred-galaxies} displays the stellar mass versus maximality for all barred galaxies in our samples. In this figure, we used the total stellar mass, which includes all the bounded star particles to each galaxy. It appears clear that high mass barred galaxies are all maximal in TNG50. The approximation used to compute the baryonic rotation curve is actually underestimating the maximality in TNG50 \citep[see, e.g.,][figure 2.17]{BT08}, so this approximation is not the reason for this result. The colour of each point indicates the bar strength for galaxies in TNG50, and the grey points correspond to the barred galaxies in the SPARC sample. Clearly, some observed galaxies, like NGC~1090, can be barred while being slightly submaximal, whilst this never happens in TNG50. Note that the error bars on the observed $\mathcal{F}_b$ are slightly underestimated since the value should never go above 1, but the difference between observations and simulations at stellar masses above $\sim 10^{11} M_\odot$ are too large to be accounted for by slightly larger error bars. Also, in order to check that the definition of the stellar mass of galaxies did not affect our results, we repeated our analysis for the stellar masses measured within a radius of $30$ kpc in simulated data. In this way, the coloured points in Fig.~\ref{fig:barred-galaxies} would be slightly shifted in the vertical direction, but this did not change the overall conclusion.

As already noted in the introduction, the lack of barred galaxies with $M_* < 10^{10.5} M_\odot$ in TNG50 was already put forward in \citet{Roshan2021}, where other problems on the too small size of bars for their pattern speed were also put forward \citep[see also][]{Frankel2022}. The problem highlighted here is not a simple rehearsal of these previously mentioned ones. Indeed, if we restrict ourselves to galaxies with $M_* \geq 10^{10.5} M_\odot$, we get 45 galaxies in the SPARC, which 15.6\% of them are submaximal. On the other hand, in TNG50, we have 230 galaxies in this mass range and only 7 (3\%) submaximal discs. We find that at least 57\% of submaximal galaxies in the SPARC are barred, while for TNG50, this fraction is 14.3\%. This means that our result is an interplay between the fact that TNG50 does not produce enough barred submaximal discs and the fact that it does not produce enough submaximal/near-maximal discs at stellar masses $M_* \geq 10^{10.5} M_\odot$, where most observed barred submaximal discs lie (see Fig.~\ref{fig:barred-galaxies}). Note that bars are also observed in low-surface brightness submaximal discs with $M_*< 10^{10} M_\odot$ \citep{Peters2019}, which were not included in the present study.



\section{Conclusion} \label{sec:conc}
We explored submaximal discs in the SPARC sample and galaxies at $z=0$ in TNG50 simulation, based on the criterion defined in Eq.~\ref{criterion}. We showed that the distribution of galaxies in terms of maximality is not so different for these data sets but that the fraction of barred submaximal discs is not the same. Focusing on galaxies with $M_{\star}\geq 10^{10} M_{\odot}$, we found that although the SPARC sample biases against the selection of barred galaxies, $36\%$ of submaximal discs in this sample are barred, but only $5.8\%$ of the submaximal discs in TNG50 host a bar and there is no strongly barred submaximal disc in this data set. Quantitatively, we showed that the probability of having the observed number of barred submaximal discs in SPARC from the TNG50 statistics is always lower than 2.5\%, even if we restrict ourselves to near-maximal discs. We point out that the problem highlighted here can be seen as an interplay between the fact that TNG50 does not produce enough barred submaximal discs and the fact that it does not produce enough submaximal/near-maximal discs at high stellar masses ($M_* \geq 10^{10.5} M_\odot$,), where many observed barred submaximal discs lie. The roles of tidal interactions with close neighbours, of the efficiency of the cooling in the gaseous discs, and even of the numerical resolution will have to be further investigated in the future to understand this discrepancy. Our results, in any case, add to the challenge of forming the right statistics and properties of barred galaxies in $\Lambda$CDM simulations of galaxy formation.
 
\section*{Acknowledgements}
This research has used the "Aladin sky atlas" developed at CDS, Strasbourg Observatory, France \citep{Bonnarel2000, Boch2014}. We would like to thank Mohammad Hosseinirad for downloading the TNG50 data. This work is supported by Ferdowsi University of Mashhad under Grant No. 56145 (13/9/1400). BF acknowledges funding from the European Research Council (ERC) under the European Unions Horizon 2020 research and innovation programme (grant agreement No. 834148) and from the Agence Nationale de la Recherche (ANR projects ANR-18-CE31-0006 and ANR-19-CE31-0017).
\section*{Data Availability}
The cosmological simulation data used in this letter are publicly available on the IllustrisTNG website\footnote{https://www.tng-project.org/data/}. The observational data set used in this letter is also available online\footnote{astroweb.cwru.edu/SPARC} (see Section \ref{sec:SPARC} for the relevant references). The data generated in this letter will be shared on reasonable request to the corresponding authors.


\bibliographystyle{mnras}
\bibliography{SPARC_TNG} 




%
%


\bsp	
\label{lastpage}
\end{document}